# Modern Physics demonstrations with DIY Smartphone Spectrometers

Aarushi Khandelwal[1], Tze Kwang Leong[2,†,] Yarong Yang[2], Loo Kang Wee[3], Félix J. García Clemente[4], T Venkatesan[1], and Hariom Jani[1*]

[1]NUS Nanoscience and Nanotechnology Initiative, National University of Singapore, Singapore
[2]Sciences Branch, Curriculum Planning and Development Division, Ministry of Education, Singapore
[3]Educational Technology Division, Ministry of Education, Singapore
[4]Department of Computer Engineering and Technology, University of Murcia, Spain
[†]leongtzekwang@gmail.com
[*]hariom.k.jani@u.nus.edu

**Abstract**

Smartphones are widely available and used extensively by students worldwide. These phones often come equipped with high-quality cameras that can be combined with basic optical elements to build a cost-effective DIY spectrometer. Here, we discuss a series of demonstrations and pedagogical exercises, accompanied by our DIY diffractive spectrometer that uses a free web platform for instant spectral analysis. Specifically, these demonstrations can be used to encourage hands-on and inquiry-based learning of wave optics, broadband vs discrete light emission, quantization, Heisenberg's energy-time uncertainty relation, and the use of spectroscopy in day-to-day life. Hence, these simple tools can be readily deployed in high school classrooms to communicate the practices of science.

## 1. Introduction

Optical spectroscopy provides a critical bridge between classical and modern physics, linking concepts of waves, optics, absorption, emission, and quantization of energy levels.[1] For high school students, it is thus an important tool to ground otherwise abstract concepts in modern physics while also demonstrating a material finger-printing technique widely used in scientific research today.

Over the recent years, the proliferation of smartphone spectrometers, which take advantage of the good quality in-built cameras and processing capabilities of smartphones, has made optical spectroscopy cheaper and more accessible[2,3,4]. Of particular interest in classroom demonstrations are DIY spectrometers[5,6,7], which are usually made of a frame (either paper or 3D printed) and a diffraction grating mounted onto the smartphone camera that converts the diffracted light to an optical spectrum. Although such DIY spectrometers may not enjoy as high resolution or signal-to-noise ratios as industry- or research-grade counterparts, they are very useful in a pedagogical setting. Students can tinker with the DIY components to understand how principles of optics can be exploited in spectroscopy and appreciate how readily available smartphones can be used to study classical and quantum phenomena.

In this paper, we present a framework for classroom demonstration exercises using our DIY smartphone diffractive spectrometer that was developed for implementation in Junior Colleges across Singapore[8,9]. The advantage of our approach is the access to a free web platform that allows instant conversion of the diffraction patterns into calibrated line spectra. This enables students to not only inspect the colors visually, but also develop a deeper appreciation of the intensity vs wavelength distributions. Our goal is to supplement the existing modern physics syllabus and facilitate student-centric learning through engaging and investigative hands-on experiences. The demonstrations were scaffolded by a predict–explain–observe–explain (PEOE) pedagogy[10,11] to better engage and guide students (see supplementary information for details on questionnaires).

The pedagogical exercise series covers in-syllabus concepts such as diffraction gratings, quantization of electronic energy levels of atoms, discrete and broadband spectra, and the use of spectrometry in gas identification. It also stretches beyond the syllabus to enhance student understanding of Heisenberg's energy-time uncertainty principle by exploring spectral pressure broadening[12]. It concludes with some phone-based demonstrations, intended to fulfill the "practices of science" objective of relating science and society by demonstrating the day-to-day implementations of physics principles.[13]

## 2. Experimental Setup

A DIY spectrometer and a smartphone are required for all demonstrations in this series. The four other components needed are (i) standard discharge tubes of gases (e.g., hydrogen, helium, cadmium, etc.), (ii) mercury high-pressure and low-pressure lamps, (iii) incandescent bulbs, and (iv) a second smartphone that will not be attached to the spectrometer. Alternatives to these components are suggested in the demonstration descriptions when possible.

In our deployment of this series, we developed the smartphone spectrometer introduced in Ref. 7, see Figure 1, which must be paired with an android phone or a laptop that can access the website in Ref. 14 to convert the first-order diffracted light ($\lambda = d \sin(\theta)$) into the final spectrum. The wavelengths detected by the setup can be calibrated with a fluorescent light source (see supplementary information), which is readily available in typical classrooms.

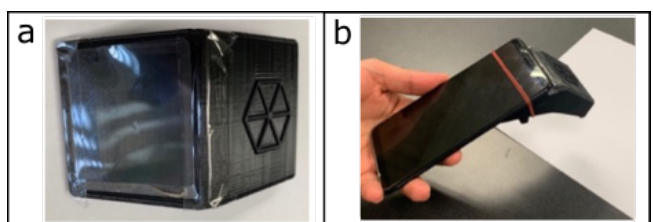

*Figure 1: Smartphone spectrometer (a) top view of the DIY spectrometer showing the diffraction grating in a 3D printed holder with the grating lines parallel to the long edge and (b) side view when performing calibration of the spectrometer attached over the camera of the smartphone with a rubber band.*

Alternatively, this series of demonstrations can also be deployed in classrooms with alternate smartphone spectrometers accompanied by stand-alone apps, like FrinGOe[3] and SpectraSnapp[15].

## 3. Understanding the smartphone spectrometer

The DIY components allow students to perform two pre-demonstration exercises studying sunlight to better understand the working of the spectrometer.

First, before the spectrometer is attached to the smartphone, students should point the spectrometer at a blank paper placed in bright sunlight and look through its clear side. They will see Figure 2(a), the image that the smartphone camera will receive and convert to a spectrum via the app. This allows them to review optics as they observe how the diffraction grating in the spectrometer splits the sunlight into different colors.

Then, after attaching the smartphone to the spectrometer, they will record the spectrum of sunlight, Figure 2(b). The deviation of this spectrum from the sun's expected black body radiation spectrum, with dips in intensities at frequencies away from the sensors' optimal frequencies, allows them to understand how the smartphone processes color input predominantly using its color sensors: red, green, and blue. Depending on the device, sensors may saturate due to varying color sensor sensitivity or intense brightness, as seen in Figure 2(b) where there is an artificial dip in the green region (500 – 600 nm).

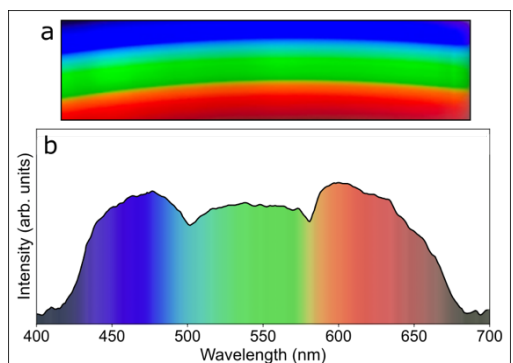

*Figure 2: Sunlight through (a) a diffraction grating and (b) the spectrometer.*

Finally, in-class discussions focusing on the practices of science can also highlight the limitations of the spectrometer's data due to the DIY nature of the equipment, including limited wavelength resolution (resulting in the inability to distinguish very close peaks), calibration errors (resulting in minor shifts in peak positions), sensor saturation (resulting in flattening of very strong peaks), and the influence of the processing environment (causing changes in the background of the collected spectra). Hence, such DIY spectrometers should be used as qualitative rather than quantitative pedagogical tools.

## 4. Demonstrations and discussion

There are four demonstrations in this series, focused on high school optics and modern physics.

### *4.1 Discrete and broadband sources*

First, students compare a broadband source, like sunlight, with a discrete source, like a discharge tube. While any broadband and discrete source would be sufficient, incandescent bulbs and sodium gas lamps are recommended because they produce similarly colored light. This helps forestall student confusion about whether the difference in the emission spectra observed comes from the color of the light or the nature of the source. The incandescent bulbs may need to be covered with plain printing paper to reduce brightness and prevent saturation of the DIY spectrometer.

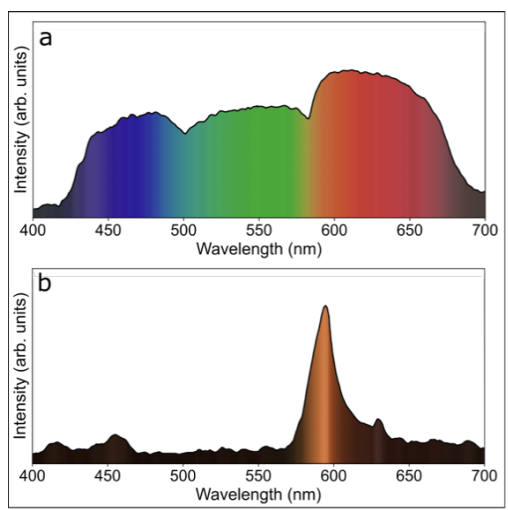

*Figure 3: Emission spectrum of (a) incandescent bulb and (b) sodium gas lamp.*

Students will observe that the incandescent bulb has a continuous spectrum, Figure 3(a), while the sodium gas lamp has a discrete spectrum with emission peaking at the sodium d-lines, Figure 3(b). Note that the d-lines looked merged due to the limited resolution.

As both light sources measured produce similarly colored yellow light, students may find the observed difference between the spectra surprising, and this should be used to motivate discussions between classical and quantum effects. The incandescent bulb is a classical broadband source. The electrical heating of the tungsten filament inside produces blackbody radiation, with wavelengths across the visible spectrum, appearing yellow overall due to the typical bulb operating temperatures.

In contrast, the sodium gas lamp has a discrete spectrum because it is a quantum source. To explain the emission spectrum, students must recall the Bohr model. Electrically heating the sodium gas excites electrons in sodium atoms to higher energy levels, and they emit light as they relax back to specific lower levels. The sodium gas lamp appears yellow because of the prominent d-lines that correspond to an electron transitioning from the 3p to 3s levels in sodium atoms.

*4.2 Unique emission spectra of elements*

Second, students can compare the emission spectra from discharge tubes of various elements. In Figure 4, we show an example of the cadmium discharge tube, which can be contrasted with the sodium gas lamp spectrum in Figure 3(b).

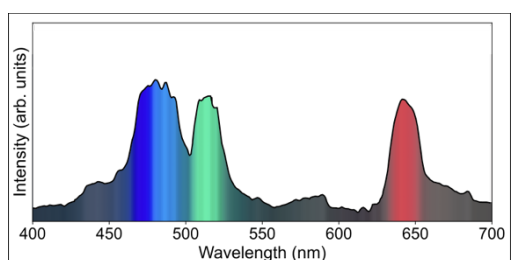

*Figure 4: Emission spectrum from discharge tube of cadmium.*

To explain the difference between the discrete emission spectra from discharge tubes of different elements, students will need to recall quantization and the Bohr model. Every element has different electronic energy levels, and thus the electronic transitions and emission spectra are unique to each element. During classroom discussion, this could also be linked to the use of optical spectroscopy for element identification, e.g., in chemistry and astrophysics.[16,17]

*4.3 Energy-time uncertainty principle*

Third, students can experimentally test the otherwise abstract concept of Heisenberg's uncertainty principle by studying the phenomenon of pressure broadening. This demonstration can be conducted with any discharge tube(s) where the gas pressure can be controlled directly. However, as standard discharge tubes do not have this feature, mercury gas lamps at different pressures can also be used.

Students observe that the emission spectrum of the high-pressure lamp, Figure 5(a), has significantly broader features than the low-pressure lamp, Figure 5(b). This is a simple demonstration of the energy-time uncertainty principle, $\Delta E \Delta t \geq \hbar/2$, which asserts that a quantum state existing for a short finite duration cannot have a definite transition energy. For instance, a higher gas pressure leads to increased collisions between the atoms in the gas. This decreases the average time between their collisions, thereby reducing the characteristic lifetime of the atoms in the excited quantum state ($\Delta t$). Per the uncertainty principle, this leads to an increase in the uncertainty of the photon energy emitted during the electronic transition ($\Delta E$), which is detected experimentally as the broadening of the emission spectral peaks.

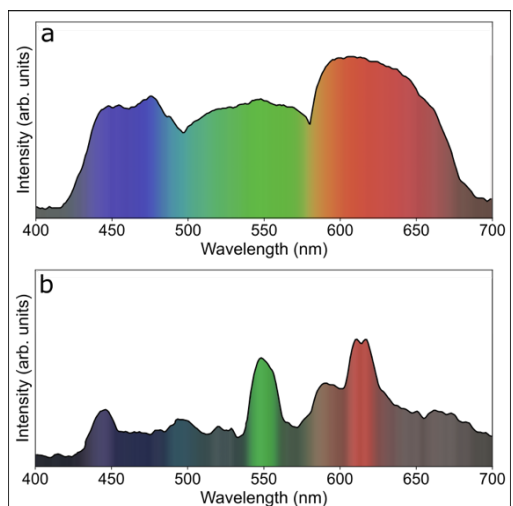

*Figure 5: Mercury emission spectrum in (a) high-pressure lamp and (b) low-pressure lamp.*

Alternative implementations of this demonstration could use high- and low-pressure lamps of a different element, like sodium, or use an air tube, where an attached vacuum pump is used to gradually vary the pressure of air in the discharge tube.

### *4.4 Phone experiments*

Finally, students conduct two phone-related experiments to understand how their devices produce color and thus relate modern physics and optics concepts to their daily lives.

First is the "color shift" experiment, which examines how the measured spectrum changes as the color of the screen changes. Students should focus the spectrometer on a second smartphone screen that plays a video cycling through various different colors (see Supplementary information for details). As the colors change, they observe the evolution of the emission spectrum. Figure 6 shows the resulting spectra at three composite colors.

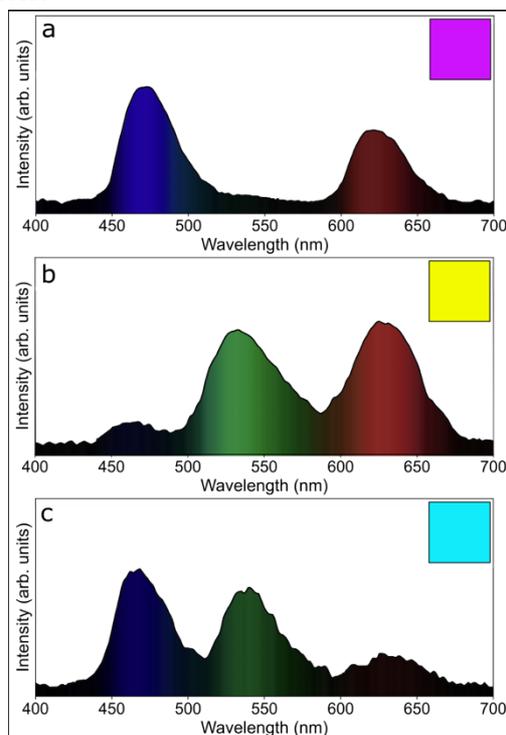

*Figure 6: Color shift experiment. Spectrum from a (a) magenta, (b) yellow, and (c) cyan smartphone screen with insets of the screen color.*

These results contrast with many students' initial prediction that they would observe a narrow peak that shifts its mean wavelength depending on the screen color. Thus, they are surprised when the spectra demonstrate that the gamut of colors on a smartphone screen is produced by tuning the intensities of the three primary color peaks at constant wavelength positions. This shows how our devices, which construct all colors from red, green, and blue colored LEDs, fundamentally differ from broadband light sources and discharge gas tubes.

Second is the "night mode" experiment, where students measure the spectrum of a white screen of the second smartphone with its night mode turned off and on, as shown in Figure 7 (see supplementary information for details).

Students will observe that the phone's night mode shifts the screen to warmer colors and causes a reduction in the blue light peak in the spectrum. This can be used to discuss the role of technology in our society, as the night mode was created after excessive blue light at night was found to disrupt the circadian rhythm and increase sleep and psychiatric disorders, obesity, diabetes, and some types of cancers.[18,19]

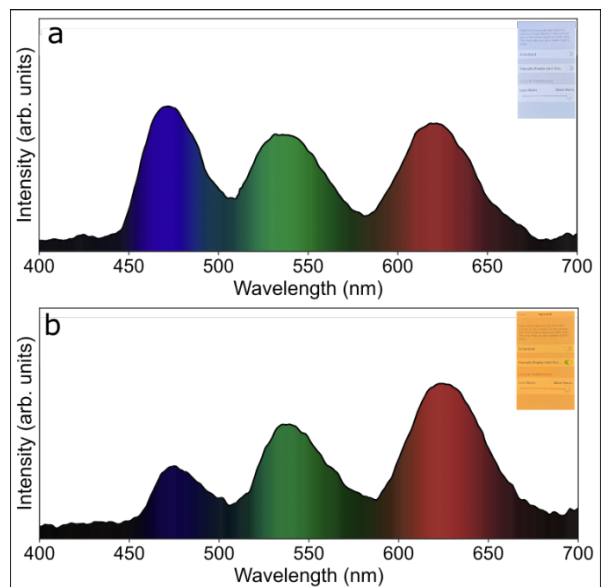

*Figure 7: Night mode experiment. Spectrum of a white section of the smartphone screen (a) without and (b) with night mode. Insets show the smartphone screen.*

## 5. Conclusion

Here, we discussed a series of demonstrations scaffolded by investigation-driven pedagogical exercises that make use of a DIY smartphone spectrometer. These exercises can help to foster an engaging hands-on and student-centric learning of the modern physics and optics curriculum in high school classrooms.

**Acknowledgments**

This MAF study was funded by the Education Research Funding Programme, National Institute of Education (NIE), Nanyang Technological University, Singapore, project number AFD 04/16 VV and the experiments were scaled up to all Junior Colleges in Singapore with the support of AST. We thank Mr. J. B. Tan, Mr. H. Tan, Dr. D Wong, and Dr. D. Tan for their support on this project. We also thank Dr. A. Patra and Dr. A. Srinivasan for discussions and support. The views expressed in this paper are the authors' and do not necessarily represent the views of the host institution.

# Supplementary information

## 1. Using the smartphone spectrometer

In our classroom demonstrations, we used the smartphone spectrometer briefly described in Ref. R1. After the spectrometer has been 3D-printed and the diffraction grating has been inserted, it must be attached to an Android Phone camera or a laptop camera with the clear side facing the camera. The spectrum is analyzed using the web platform[R2] that can be opened on a browser in the phone spectrometer.

Before the spectrometer can be used, its wavelength range should be calibrated with a fluorescent light source. This is to convert the angular separation of the first order diffracted light from the 1000 lines/mm grating into corresponding wavelengths using the relation $n\lambda = d\sin(\theta)$. This can be done by (i) pointing the spectrometer to a white sheet of paper illuminated by fluorescent lights in the room, and (ii) pressing the "Calibrate" button on the website. Once the calibration is complete the relative positioning of the phone and spectrometer should not be changed, or else the setup will need to be re-calibrated.

The spectra can then be measured by pointing the spectrometer at the desired source, keeping the phone horizontal, ensuring the spectral lines overlap the long, horizontal, dotted line across the screen[R2], and clicking "measure". The phone flash must be off while the measurements are collected. If the signal is too saturated, the spectrometer should be moved further away from the source or the light should be measured off a piece of white paper. Saturation of sensors at high intensity peaks can cause apparent splitting or suppression of single peaks.

## 2. Color shift video

The video cycling through various colors necessary for the color shift experiment can be accessed using the link in Ref. R3. It should be played in full screen so the DIY spectrometer can be pointed anywhere on the screen.

## 3. Accessing night mode on different phones

On iPhones, the night shift can be enabled by going to Settings > Display & Brightness > Night shift and turning on Manually Enable Until Tomorrow. The "color temperature" can also be adjusted, where increasing warmth would lead to a greater reduction in blue light when the night shift is on.[R4]

In some Androids, the night shift can be enabled by either going to Settings > Display > Night Light or Settings > Display > Blue Light Feature, depending on the phone. If this is not available, there are several third-party apps that can be installed to achieve the night shift, depending on the phone's make and model.

## 4. Safety precautions

To prevent accidents, the following safety precautions are required when handling the high-power discharge tubes or gas lamps:

1. Be mindful of the allowed voltage recommendation. Do not touch any wires directly as a very high voltage runs through the system.
2. Turn off the voltage immediately if there are strange smells, flickering lights, or burning inside the tube.
3. Handle the tubes with care as they are made of glass and can break into shards.
4. They can get very hot and should not be touched directly until they cool down after being switched off to avoid burns. Wait for at least 5 mins before changing the tube after an operation.
5. Some discharge tubes contain elements like Mercury or Hydrogen, which are hazardous, and should be disposed of carefully in case of any cracks or leaks.

*5. Introduction for intellectual scaffolding*

Sufficient intellectual scaffolding provided before this series of demonstrations is important for students to grasp the key learning objectives[R5,R6]. Specifically, familiarity with the following concepts would be beneficial:

- Diffraction gratings and optical interference
- The Bohr model of the atom arising from the wave nature of electrons
- Discrete emission spectra arising from quantized energy levels in atoms
- The link between light color and wavelength
- Broadband blackbody radiation spectra
- Heisenberg's uncertainty principles

*6. PEOE Questionnaires*

A set of 5 questionnaires can be given to the students to stimulate their thinking and deepen their understanding of the demonstrations based on the inquiry-driven Predict-Explain-Observe-Explain (PEOE) framework. They are included here as a reference:

- Pre-demo questions: *To help clear conceptual issues before the experiment*
- Demonstration questionnaire (Discrete and broadband sources)
- Demonstration questionnaire (Energy-time uncertainty principle)
- Demonstration questionnaire (Phone experiments)
- Post-demo questions: *To help evaluate the level of understanding after the experiment*

Our model answers to the MCQ questions are highlighted below.

**Questionnaire 1: Pre-demo questions**

1. What is the origin of the emission spectra observed from a gas source?
   A. Relaxation of an electron from high energy to low energy
   B. Relaxation of a proton from high energy to low energy
   C. Excitation of an electron from low energy to high energy
   D. Absorption of light

2. When a discharge tube is set up, the intensity vs wavelength graph of the emission spectra of the gas inside over the whole electromagnetic spectrum: (choose all that apply)
   A. Can have a single peak
   B. Can have multiple peaks
   C. Is dependent on the gas composition
   D. Is dependent on the pressure of the gas
   E. A and C only

3. When the gas in the discharge tube is changed from hydrogen to neon:
   A. The peak positions remain the same, but their intensities increase
   B. The peak positions are shifted proportionally to the change in average mass of the gaseous species
   C. The peaks and their intensities remain the same
   D. The peaks of the two graphs are not related to each other

4. When pressure is increased, what happens to the collision time of the molecules of gas in the discharge tube?
   A. Increases
   B. Decreases
   C. Stays the same
   D. Depends on the gas

5. If the spacing between lines in the diffraction grating used in the smartphone spectrometer is decreased (i.e. density of lines is increased), yet the diffracted beams can still be fully captured by the camera aperture, then the resolution of the spectrometer would:
   A. Increase
   B. Decrease
   C. Stay the same

6. A common smartphone screen has over a million subpixels that it uses to generate images. Which of the following sentences are correct?
   A. There are Red, Orange, Yellow, Green, Blue, Indigo, and Violet colored subpixels
   B. There are only Red, Green, and Blue subpixels
   C. There are only Red, Yellow, and Blue subpixels
   D. Each subpixel can change color according to the image required

**Questionnaire 2: Demonstration questionnaire (Discrete and broadband sources)**
*Measure the spectra from an incandescent bulb and a sodium gas lamp.*

*Predict and Explain Questions (before carrying out the experiment)*
1. Sketch the intensity vs wavelength spectrum you see expect to see from an incandescent bulb.
2. A sodium gas lamp has a similar color to the incandescent bulb. Sketch the intensity vs wavelength spectrum you see expect to see from a Sodium gas lamp.
3. Are your predictions for the bulb and gas lamp spectra different? Explain why/why not?

*Observe and Explain Questions (after carrying out the experiment)*
4. Sketch the intensity vs wavelength spectrum you observe from the incandescent bulb.
5. Does the spectrum match your prediction? Explain why/why not?
6. Sketch the intensity vs wavelength spectrum you observe from a sodium gas lamp.
7. Does the spectrum match your prediction? Explain why/why not?
8. Why are the spectra from the incandescent bulb and gas lamp different, despite having a similar color?

**Questionnaire 3: Demonstration questionnaire (Energy-time uncertainty principle)**
*Measure the spectra from high-pressure and low-pressure Mercury lamps.*

*Predict and Explain Questions (before carrying out the experiment)*
1. In this experiment, you are considering the spectra from a high-pressure and low-pressure Mercury lamp. Do you expect to see any difference in their spectra? If so, what kind of difference and why do you expect to see this?

*Observe and Explain Questions (after carrying out the experiment)*
2. Sketch the intensity vs wavelength spectrum you observe from the high-pressure mercury lamp.
3. Sketch the intensity vs wavelength spectrum you observe from the low-pressure mercury lamp.
4. Explain how and why the width of the peaks change as the pressure is increased. (Hint: use Heisenberg's Energy-Time uncertainty principle).

**Questionnaire 4: Demonstration questionnaire (Phone experiments)**
*In this experiment, you will measure spectra from the screen of a second phone.*

*Color Shift Predict and Explain Questions (before carrying out the experiment)*
Below is the visible light spectrum.

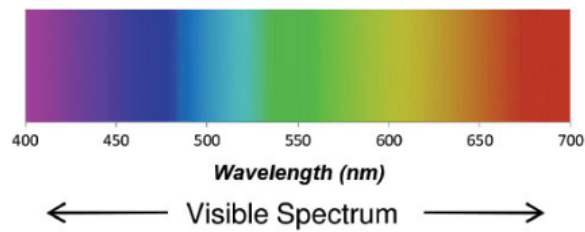

1. When the color of the phone screen changes (for example, from purple to orange), predict how you expect the intensity vs wavelength spectrum measured from the phone screen to change?

*Color Shift Observe and Explain Questions (after carrying out the experiment)*
2. Sketch the spectrum of intensity vs wavelength you see when the light is purple.
3. Sketch the spectrum of intensity vs wavelength you see when the light is orange.
4. Does the spectra change with color as you predicted? Explain why or why not?

*Night Shift Predict and Explain Questions (before carrying out the experiment)*
5. Predict how the spectrum measured from a white phone screen changes when the night shift is turned on.

*Night Shift Observe and Explain Questions (after carrying out the experiment)*
6. Sketch the spectrum of intensity vs wavelength before the night shift.
7. Sketch the spectrum of intensity vs wavelength after the night shift.
8. Does the difference between these spectra match your predictions? Explain why or why not.

**Questionnaire 5: Post-demo questions**
1. The intensity vs wavelength graph of the emission spectra of a gas inside a discharge tube over the whole electromagnetic spectrum: (choose all that apply)
    A. Must have only a single peak
    B. Is dependent on the gas
    C. Is independent of temperature
    D. Is independent of pressure
    E. None of the above

2. Are the position, shape, and intensity of the spectral peaks only dependent on the color?
    A. Yes, it only depends on the color
    B. No, it also depends on the source

3. When pressure is reduced, what happens to the collision time of the molecules of gas in the discharge tube?
   A. **Increases**
   B. Decreases
   C. Remains the same
   D. Depends on the gas

4. When temperature is increased, what happens to the collision time of the molecules of gas in the discharge tube?
   A. Increases
   B. **Decreases**
   C. Remains the same
   D. Depends on the gas

5. If the width of the spectral peaks measured narrows due to a change in the pressure, which of the following are true?
   A. Energy uncertainty increased
   B. **Energy uncertainty decreased**
   C. Energy uncertainty remains the same

6. If the width of the spectral peaks measured narrows due to a change in the pressure, which of the following are true?
   A. **Time between collisions increases**
   B. Time between collisions decreases
   C. Time between collisions remains the same

7. The night shift reduces the blue light emitted by the smartphone. If one wanted to reduce the yellow light emitted instead, they would need to:
   A. Reduce the intensity of light emitted by red and blue subpixels
   B. Reduce the intensity of light emitted by yellow subpixels
   C. **Reduce the intensity of light emitted by red and green subpixels**
   D. Reduce the intensity of light emitted by green and blue subpixels

8. Which of the following would improve the resolution of the spectrometer (ability of the spectra to separate closely positioned peaks)? (Check all that apply):
   A. **Look at a higher order diffraction (e.g. n = 2)**
   B. Look at the zeroth order diffraction
   C. **Increase the distance between the diffraction grating and the smartphone camera**

9. What are possible sources of error in the spectra observed by the DIY spectrometer?
   A. Saturation of smartphone camera sensors near very intense spectral peaks
   B. Shift in the spectral background during measurement if other light sources are present nearby
   C. Movement of the relative position of the smartphone and DIY spectrometer can cause some shift in spectral positions
   D. **All of the above**